\documentclass[11pt]{article}

\usepackage{amsmath}
\usepackage{color}
\usepackage{graphicx}
\usepackage{amssymb}
\usepackage{bbm} 
\usepackage{float}
\usepackage{scrtime}
\usepackage{fancyhdr}
\usepackage{subfigure}
\usepackage{authblk}
\usepackage{lipsum}
\usepackage{rotating}
\usepackage{floatpag}
\usepackage{varioref}
\usepackage{slashed}
\usepackage{enumerate}

\usepackage[colorlinks=true
,urlcolor=blue
,anchorcolor=blue
,citecolor=blue
,filecolor=blue
,linkcolor=black
,menucolor=blue
,pagecolor=blue
,linktocpage=true
]{hyperref}

\newcommand{\be}{\begin{equation}}
\newcommand{\ee}{\end{equation}}
\newcommand{\bea}{\begin{eqnarray}}
\newcommand{\eea}{\end{eqnarray}}

\newcommand{\fk}{\ensuremath{\mathcal{K}}}

\newcommand{\fw}{\ensuremath{\mathcal{W}}}

\begin{document}

\floatpagestyle{plain}

\pagenumbering{roman}

\renewcommand{\headrulewidth}{0pt}
\rhead{
OHSTPY-HEP-T-14-003
}
\fancyfoot{}

\title{\huge \bf{Chaotic Hybrid Inflation with a Gauged B - L}}

\author{Linda M. Carpenter}
\author{Stuart Raby}

\affil{\em Department of Physics, The Ohio State University,\newline
191 W.~Woodruff Ave, Columbus, OH 43210, USA \enspace\enspace\enspace\enspace \medskip}

\maketitle
\thispagestyle{fancy}

\begin{abstract}\normalsize\parindent 0pt\parskip 5pt
In this paper we present a novel formulation of chaotic hybrid inflation in supergravity.   The model includes
a waterfall field which spontaneously breaks a gauged $U_1(B-L)$ at a GUT scale.  This allows for the possibility of future
model building which includes the standard formulation of baryogenesis via leptogenesis with the waterfall field decaying into
right-handed neutrinos.  We have not considered the following issues in this short paper,i.e.  supersymmetry breaking, dark matter or the gravitino or moduli problems.   Our focus is on showing the compatibility of the present model with Planck, WMAP and Bicep2 data.
\end{abstract}

\clearpage
\newpage

\pagenumbering{arabic}

\section{Introduction}

The recent measurement of B modes by the Bicep2 collaboration \cite{Ade:2014xna} has generated great interest,
since it is possibly the first direct measurement of quantum gravitational excitations resulting from a period of
inflation in the early universe.   The tensor to scalar ratio was measured to be $r = 0.16^{+0.06}_{-0.05}$ with $r = 0$ disfavored at 5.9 $\sigma$. Combined with the Planck plus WMAP measurement of scalar density perturbations for the $\Lambda$CDM model, $3.089^{+0.024}_{-0.027} \times 10^{-10}$ \cite{Ade:2013uln}, gives a value of the inflaton potential density at inflation, $V = (2.2 \times 10^{16} \ {\rm GeV})^4 \frac{r}{0.2}$.
This value is tantalizingly close to the supersymmetric grand unified [SUSY GUT] scale to suggest some connection to SUSY GUTs.  In this paper
we combine chaotic inflation with a waterfall field which at the end of inflation spontaneously breaks a conserved gauged $B - L$ symmetry, i.e. the first step towards a SUSY GUT.  We refer to this as chaotic hybrid inflation, even though there is only one inflaton.  Note, current hybrid inflation models are small field models which are inconsistent with the BICEP 2 measurement (see for example \cite{Buchmuller:2012wn,Buchmuller:2013dja}).    The parameters of our model need not be finely tuned to produce this behavior.  Our model represents an existence proof for a class of such simple models. We also show how to incorporate leptogenesis in this analysis.   Similar work in this direction which however does not include the waterfall field has recently appeared in the literature \cite{Murayama:2014saa}, see also earlier papers on chaotic inflation in supergravity \cite{Linde:1983gd,Goncharov:1983mw,Linde:1986fc,Kawasaki:2000yn,Kawasaki:2000ws,Yamaguchi:2003fp}.

\section{Model Building Checklist}

 In order to exhibit both chaotic inflation, and the activation of waterfall fields at the end of the inflationary trajectory, we require a model with the following properties: \\

$\bullet$ There must exist an inflation field, exhibiting a shift symmetry, which exhibits a slow roll from a trans-Plankian vev down to small field values.

$\bullet$ The inflaton must get a mass such that during inflation its energy density is the GUT scale.

$\bullet$ All other fields in the theory must not be tachyonic during the inflationary trajectory.

$\bullet$ Waterfall fields must exist in the model, and must exhibit positive quadratic potentials during inflation.

$\bullet$ As the inflaton field nears the end of its inflationary trajectory, the waterfall potential must exhibit a saddle point at zero field value and fall into a minimum with nonzero expectation value. \\

We also add some further requirements on potential models in order to avoid other observational pitfalls: \\

$\bullet$ Any field which couples to the inflaton must have a mass larger than the Hubble scale H,  to avoid producing iso-curvature perturbations of the metric. \\

$\bullet$ Supersymmetry must be unbroken at the end of inflation. \\

In the next section we present a model with a quadratic potential for the inflaton field, $\Phi$, and waterfall fields, $S_{1,2}$.
The waterfall fields have non-zero $B-L$ charge.

\section{The Inflaton Sector}

\begin{figure}[p]
\thisfloatpagestyle{empty}
\centering
\includegraphics[width=0.6\textwidth]{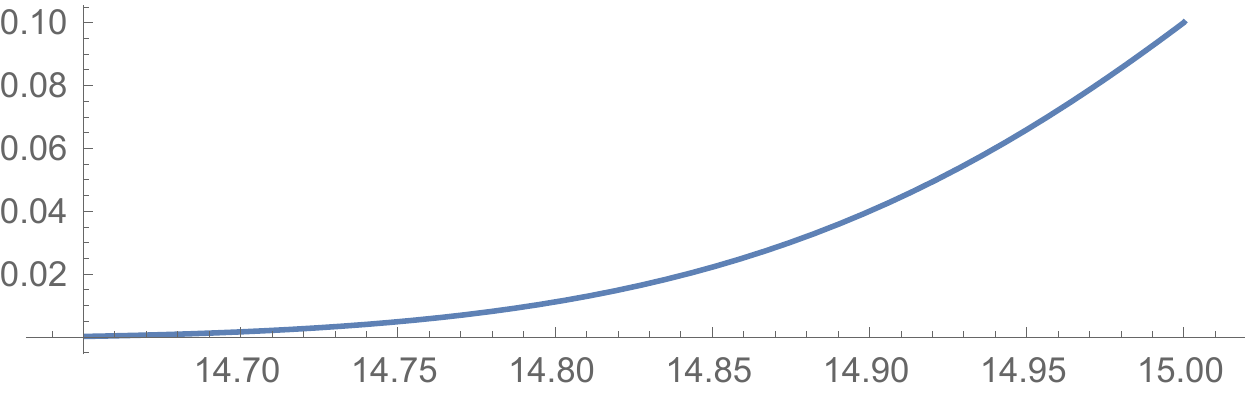}
\caption{\footnotesize{This is the numerical solution to the equations of motion for the fields, $\Phi_i, X_i$, obtained using Mathematica. The vertical (horizontal) axes are $X_i$ ($\Phi_i$) starting with initial conditions at $t=0$, $X_i[0]=0.1, \frac{d X_i}{dt}[0] = 0, \Phi_i = 15, \frac{d \Phi_i}{dt}[0] = 0$ [in Planck units].  We find $N_e = 58.3$.
}}
\label{fig:inflaton_dynamics}
\end{figure}

The inflation sector of the model includes the inflaton field, $\Phi$, the auxiliary fields, $X, \ Y$, and the
$B - L$ breaking fields, $S_1,  S_2$, which are the waterfall fields associated with the end of inflation and reheating.
The superpotential and Kahler potential for this sector is given by
\be \fw_I = \lambda \ X ( S_1 S_2 - \frac{v^2_{B-L}}{2})  +  \Phi (\kappa \ S_1 \ S_2 + m \ Y ) \ee  and
\be \fk_I =  \frac{1}{2} (\Phi + \bar \Phi)^2 + \bar X X + \bar Y Y (1 - c_Y \bar Y Y + a_Y (\bar Y Y)^2) + \bar S_1 e^{2 V_{B-L}} S_1 + \bar S_2 e^{- 2 V_{B-L}} S_2  \ee where $V_{B-L}$ is the $B-L$ gauge superfield and the $B - L$ charges of the fields $ S_1, S_2$ are $+1, -1$, respectively.\footnote{The constant $c_Y > 1/3$ in the Kahler potential is necessary for the $Y$ mass to be larger than the Hubble parameter during inflation, so that during inflation $\langle Y \rangle = 0$, and $a_Y$ is necessary for the potential to be bounded from below.  The field $X$ obtains mass during inflation proportional to $\langle \Phi \rangle$.  In addition, by adding a quartic and sextic term in the Kahler potential for $X$ it would also have a non-zero mass when $\Phi = 0$ without changing any results of the theory.}
The scalar potential is then given by
\be   V =  e^\fk (\sum_{I}\ \sum_{J} \left[ F_{\Phi_I}\ \overline{F_{\Phi_J}}\ \fk^{-1}_{I, J} - 3 |W|^2 \right]) +
\frac{D_{B-L}^2}{(S + \bar S)} + \Delta V_{CW}[\Phi_I,
\overline{\Phi}_I] \ee where $\Phi_{I} = \{ \Phi, X, S_{1,2}, \cdots \}$ and $F_{\Phi_I} \equiv \partial_{\Phi_I} \fw +
(\partial_{\Phi_I} \fk) \fw$.  The first two terms are the tree level supergravity potential and $D$ term
contribution with  $D_{B-L} = g \sum_I ( (B-L)_I \bar \phi_I \phi_I )$ where $S$ is the dilaton field.  The last term is a one loop correction which affects the vacuum energy and is negligible in our case.  The complete superpotential and Kahler potentials are given by $\fw = \fw_I + \fw_{MSSM}$ and $\fk = \fk_I + \fk_{MSSM}$ where the minimal supersymmetric standard model [MSSM] contribution is discussed in the next section.  This form of the superpotential for chaotic hybrid inflation is similar to that found in Ref. \cite{Yamaguchi:2003fp}.\footnote{Aside : an alternative form for the inflaton sector of the theory is given by \be \fw_I = \lambda \ X ( S_1 S_2 - \frac{v^2_{B-L}}{2})  +  \Phi (\kappa \ S_1 \ S_2 + \lambda^\prime \ Y \ Z) + U (Z^2 - \frac{M^2}{2}) \nonumber \ee  with $\lambda^\prime \sim 10^{-5}$ and $M \sim M_{Pl}$  and
\be \fk_I =  \frac{1}{2} (\Phi + \bar \Phi)^2 + \bar X X + \bar Y Y + \bar Z Z + \bar U U(1 - c_U \bar U U + a_U (\bar U U)^2) + \bar S_1 e^{2 V_{B-L}} S_1 + \bar S_2 e^{- 2 V_{B-L}} S_2 . \nonumber \ee}

Note, that the Kahler potential has a shift symmetry \cite{Kawasaki:2000yn} with $\Phi \rightarrow \Phi + i C$ which allows $\Phi_i \equiv Im(\Phi)$ to be super-Planckian during inflation.  Moreover the shift symmetry is softly broken in the superpotential.  In addition, the mass of the waterfall fields are of order $\kappa \langle \Phi \rangle$ during inflation.\footnote{The term proportional to $\kappa$ in the superpotential guarantees that the waterfall fields remain at zero VEV during inflation; thus allowing for efficient non-thermal leptogenesis after inflation.}  Supersymmetry is broken with scalar mass eigenvalues, in the global SUSY limit, given by $m^2_{S_\pm} = \kappa^2 |\langle \Phi \rangle|^2 \pm \lambda^2 v^2_{B-L}$ and the fermion mass eigenvalue given by $m_{S_f} = \kappa \langle \Phi \rangle$.  We take $v_{B-L}\sim 10^{-2} M_{Pl} \approx M_{GUT}$ with $M_{Pl} = 2.4 \times 10^{18}$ GeV, $\lambda = 0.1, \ \kappa = 0.01 $ and $m = 1.7 \times 10^{-5} M_{Pl} = 4 \times 10^{13}$ GeV in the following.\footnote{There is some arbitrariness in the choice of values of $\lambda, \ \kappa, \; {\rm and} \; v_{B-L}$.  These parameters, however, do not affect the inflaton trajectory during inflation.   $m$ is chosen such that the inflaton energy density during inflation satisfies, $[V(\Phi_*)]^{1/4} = M_{GUT} = 2 \times 10^{16}$ GeV.}

\begin{figure}[p]
\thisfloatpagestyle{empty}
\centering
\includegraphics[width=0.9\textwidth]{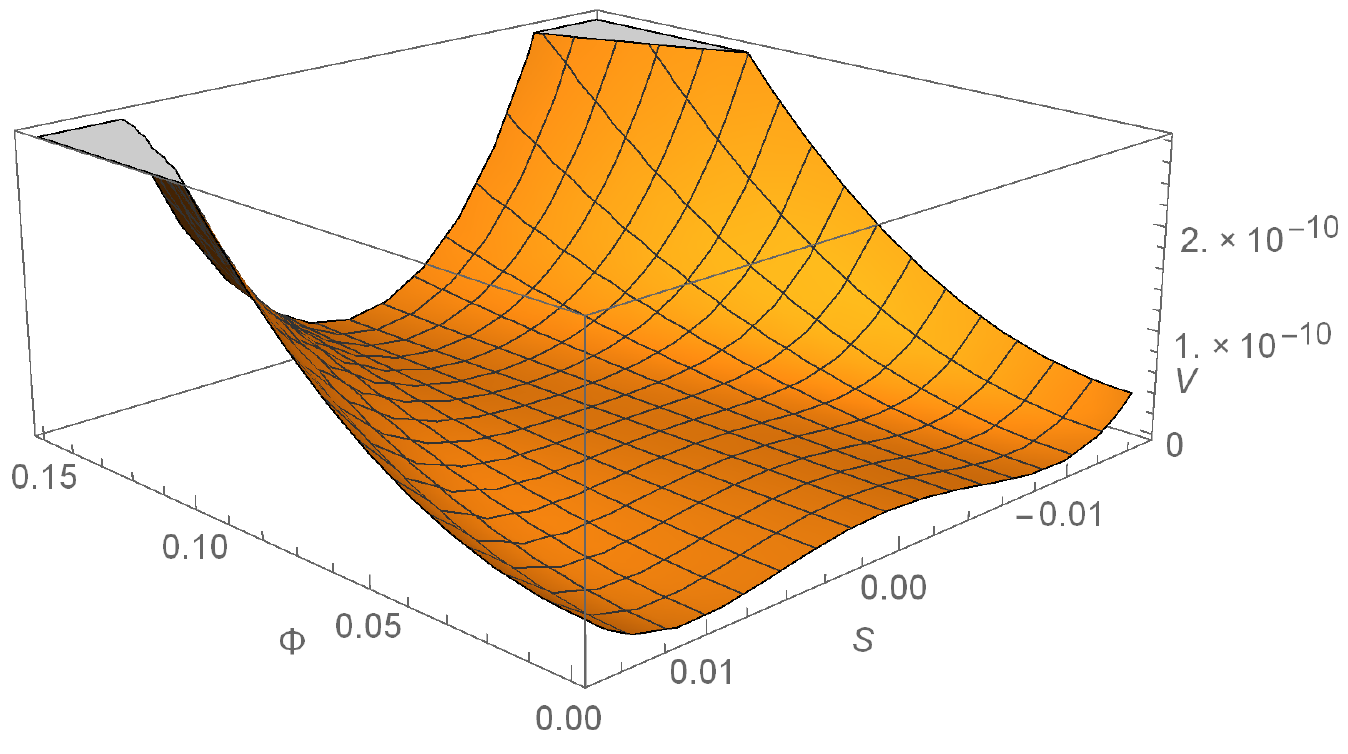}
\caption{\footnotesize{This is a 3D plot of the potential, $V$, as a function of the inflaton field, $\Phi_i$, and the waterfall field, $S_r \equiv Re(S)$,
where the field $S$ is explicitly defined in the next section.
}}
\label{fig:inflaton_waterfall}
\end{figure}
\begin{figure}[p]
\thisfloatpagestyle{empty}
\centering
\includegraphics[width=0.8\textwidth]{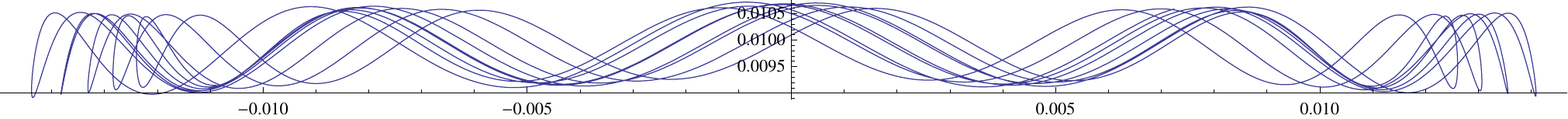}
\caption{\footnotesize{This is the numerical solution to the equations of motion for the fields, $\Phi_i, S_r$, obtained using Mathematica. The vertical (horizontal) axes are $S_r$ ($\Phi_i$) starting with initial conditions at $t=0$, $S_r[0] \equiv Re(S) = 0, \frac{d S_r}{dt}[0] = 0, \Phi_i = 15, \frac{d \Phi_i}{dt}[0] = 0$ [in Planck units].  We have only plotted the fields at the end of inflation where the waterfall field falls into its minimum.
}}
\label{fig:inflaton_dynamicsv3}
\end{figure}

The fields $S_1, \ S_2, \ X$ obtain mass proportional $\Phi_i$ during inflation.  Their minima are located at $S_1 =S_2 = X = 0$.
The slow roll parameters, $\epsilon, \eta$,  are given by \be \epsilon(\Phi_i) = \frac{1}{2} ( \frac{M_{Pl} V^\prime}{V})^2 ; \;\;  \eta(\Phi_i) = \frac{M_{Pl}^2 V^{\prime \prime}}{V} \ee with $V^\prime \equiv \frac{\partial V}{\partial \Phi_i}$ and  $V^{\prime \prime} \equiv \frac{\partial^2 V}{\partial \Phi_i^2}$.  The number of e-foldings is then given by \be N_e = \int_{\Phi_{final}}^{\Phi_*} \frac{d \Phi_i}{\sqrt{2 \epsilon}} .\ee   Slow roll ends when either $\epsilon$ or $\eta$ becomes of order one.    We can calculate analytically the slow roll parameters along the trajectory with $X = 0$.  We find $\epsilon, \eta \leq 1$ for $\Phi_i \geq 0.9 M_{Pl}$.  Starting again at $\Phi_i = \Phi_* = 15 M_{Pl}$ we find $N_e = 57.8$.  Now we can calculate the cosmological observables, the tensor to scalar ratio $r = 16 \ \epsilon(\Phi_*) = 0.14$ and the spectral index, $n_S = 1 - 6 \epsilon(\Phi_*) + 2 \eta(\Phi_*) = 0.96$.  These are all consistent with Bicep2 \cite{Ade:2014xna}  and Planck data \cite{Ade:2013uln}.

We have also checked to see how sensitive the results are to the initial conditions. We have numerically integrated the coupled field equations of motion for $\Phi_i, X_i$ (where $X_i \equiv Im(X)$).  In Figure \ref{fig:inflaton_dynamics} we plot the numerical solution to the equations of motion for the fields, $\Phi_i, X_i$, obtained using Mathematica. The vertical (horizontal) axes are $X_i$ ($\Phi_i$) starting with initial conditions at $t=0$, $X_i[0]=0.1, \frac{d X_i}{dt}[0] = 0, \Phi_i = 15, \frac{d \Phi_i}{dt}[0] = 0$ [in Planck units]. The Hubble parameter during inflation is given by $H_* = \sqrt{\frac{V(\Phi_* = 15 \ M_{Pl})}{3 M_{Pl}^2}} = 10^{14}$ GeV, corresponding to $V(\Phi_* = 15 \ M_{Pl}) \approx M_{GUT}^4$, consistent with the Bicep2 results \cite{Ade:2014xna}.  The number of e-foldings obtained with this solution given by $N_e = \int_{t_i}^{t_f} \ H[t] \ dt$ is $N_e = 58.3$.\footnote{Note, the mass of the field $X$ is given by $\Phi_i$ during inflation,  while the mass of the field $Y$ is determined by the quartic term in the Kahler potential.   Thus these fields quickly obtain their inflation era values due to Hubble friction. This is clearly seen in Figure \ref{fig:inflaton_dynamics}.}

Inflation ends when $\Phi_i \leq 0.7 M_{Pl}$ and then $\Phi_i, \ X_i$ oscillate about $\Phi_i = X_i = 0$.  As they oscillate, the energy density in the $\Phi_i, \ X_i$ fields decrease as pressureless matter until $\Phi_i \sim v_{B-L}, \ X_i = 0$ at which point the waterfall fields begin to oscillate around the supersymmetric vacuum state with $|S_1| = |S_2| =  \frac{v_{B-L}}{\sqrt{2}}$.  See Fig. \ref{fig:inflaton_waterfall} and the next section for more details on the waterfall dynamics.

\section{Coupling the Inflaton Sector to the MSSM Sector}

The MSSM includes the following fields   \be  q_i, \, u^c_i,\,  d^c_i, \,   \ell_i,\, \nu^c_i,\, e^c_i,\, H_u,\,  H_d \ee  where $ i = 1,2,3 $ is a family index.  The $B - L$ charges of the MSSM superfields are given by  \be \nu^c, +1;\;  e^c, +1;\; \ell, -1;\;   q, +1/3;\;  u^c, -1/3;\; d^c, -1/3;\; H_u, 0;\; H_d, 0 . \ee  We assume canonical Kahler potential for all MSSM fields, i.e.  \be \fk_{MSSM} =  \sum_i [ \overline{\Phi_i} e^{2 (B-L)_i V_{B-L}} \Phi_i ] .  \ee  The superpotential is given by \be \fw_{MSSM} =   \frac{\lambda_i^2}{2 M_i} (S_2 \nu^c_i)^2 + \lambda^\nu_{i j} \ell_i H_u \nu^c_j + \lambda^e_{i j} \ell_i H_d e^c_j + \lambda^u_{i j} q_i H_u u^c_j + \lambda^d_{i j} q_i H_d d^c_j - \mu H_u H_d. \ee

Following Buchm\"{u}ller et al. \cite{Buchmuller:2012wn} we can define $S_1 = S e^{i T}, \ S_2 = S e^{-i T}$.\footnote{Our fields $S_1, \ S_2$ have different $U_1(B-L)$ charges than in Ref. \cite{Buchmuller:2012wn}.  As a result the coupling to the left-handed anti-neutrinos is somewhat different.}  The complex superfield $T$ contains the Nambu-Goldstone boson of spontaneously broken $U_1(B-L)$.  In the unitary gauge $T$ is gauged away.  Then the waterfall fields $S \supset \{ s = \frac{1}{\sqrt{2}} (\sigma + i \tau), \phi; \; \psi = (\tilde s, \tilde \Phi) \}$, i.e. the scalar-Higgs and fermionic-Higgsino components respectively, decay into the left-handed anti-neutrinos, $\nu^c_i$, via the interaction in the superpotential - \be \frac{\lambda_i^2}{2 M_i} (\frac{ \sigma + i  \tau + v_{B-L}}{\sqrt{2}})^2 \nu^c_i \nu^c_i = \frac{1}{2} M_R^i \nu^c_i \nu^c_i +  \frac{h_i}{2} ( \sigma + i  \tau) \nu^c_i \nu^c_i \ee plus terms quadratic in $ S$ with $M_R^i \equiv \frac{(\lambda_i v_{B-L})^2}{2 M_i}$ and $h_i = \frac{\lambda_i^2 v_{B-L}}{M_i}$.   The fields $ \sigma, \  \tau$ are the scalar and pseudo-scalar components of $s$.

At the end of inflation with $\Phi_i \ll M_{Pl}$ the waterfall field $S_-$ develops a negative mass squared and begins to oscillate around the minimum of the potential with $\langle S_1 \rangle = \langle S_2 \rangle = \frac{v_{B-L}}{\sqrt{2}}$ with the first equality determined by the $B-L$ D-term constraint.\footnote{We assume that the dilaton VEV has been fixed in the SUSY and moduli stabilization sector of the theory. In addition, at the end of inflation, the vacuum of the waterfall sector is supersymmetric.} In Figure \ref{fig:inflaton_dynamicsv3} we plot the numerical solution to the equations of motion for the fields, $\Phi_i, S_r \equiv Re(S)$, obtained using Mathematica. The vertical (horizontal) axes are $S_r$ ($\Phi_i$) starting with initial conditions at $t=0$, $S_r[0]=0, \frac{d S_r}{dt}[0] = 0, \Phi_i = 15, \frac{d \Phi_i}{dt}[0] = 0$ [in Planck units].  We clearly see that at the end of inflation, the waterfall field goes to its minimum at $S_r = \frac{v_{B-L}}{\sqrt{2}}$.

We can now describe reheating making full use of the analysis by Buchm\"{u}ller et al. \cite{Buchmuller:2012wn,Buchmuller:2013dja}, since from this time forth the consequences of our model are identical to that described by these authors.\footnote{We only provide a simplified discussion of reheating in this paper. In particular we ignore the effects of pre-heating.  We save a more complete discussion to a future paper.}
The waterfall fields continue to oscillate around the minimum as the universe expands.  Initially, the Hubble parameter is much larger than the decay rate of the waterfall field, $S$.  The $U_1(B-L)$ gauge sector has mass of order $M_G$ and rapidly decays.  On the other hand the Higgs sector, $S$, is much lighter and decays much later.  Reheating occurs at this latter time.  When the Hubble parameter, $H = \sqrt{\frac{\rho_{S}}{3 M_{Pl}^2}}$, becomes of order the decay rate into the lightest left-handed neutrino, \bea \Gamma_{\tau \rightarrow \nu^c_1 \ \nu^c_1} \; = \; \Gamma_{\psi \rightarrow \nu^c_1 \ \tilde {\nu^c_1}^*} \; = \;  \Gamma_{\phi \rightarrow {\tilde \nu^c_1} \ {\tilde \nu^c_1}} \; = & \frac{h_1^2}{32 \pi} m_{S} (1 - \frac{4 (M_R^1)^2}{m_{ S}^2})^{1/2}, & \nonumber \\ \Gamma_{\sigma \rightarrow \nu^c_1 \ \nu^c_1}  = & \frac{h_1^2}{32 \pi} m_{ S} (1 - \frac{4 (M_R^1)^2}{m_{ S}^2})^{3/2}, & \eea with $m_{S} = \sqrt{2} e^{v_{B-L}^2/2} \lambda v_{B-L} \sim 7 \times 10^{-6} M_{Pl} \approx 1.7\times 10^{13}$ GeV, then $S$ decays and the universe reheats to $T_{RH} \approx  (\frac{90}{g^* \pi^2 } \Gamma_{\sigma}^2 M_{Pl}^2)^{1/4}$ with $g^* \sim 200$.  Given $h_1 = \frac{2 M_R^1}{v_{B-L}}$ with $M_R^1 = 5.4 \times 10^{10}$ GeV we obtain $\Gamma_{\sigma} \approx 3.42$ GeV and $T_{RH} \approx 1.3 \times 10^9$ GeV.

Given a model of neutrino masses and mixing we could then calculate leptogenesis and the resulting baryon asymmetry (see for example, Refs. \cite{Lazarides:1991wu,Asaka:1999yd,Asaka:1999jb,Buchmuller:2004nz,Buchmuller:2012wn}).  Finally supersymmetry breaking can occur in a separate sector of the theory.   For example, using the Kallosh-Linde SUSY breaking sector \cite{Kallosh:2011qk} the SUSY breaking minimum during inflation is stabilized while still having the gravitino mass of order the TeV scale.   Cosmic strings, SUSY breaking, dark matter and a solution to the strong CP problem is beyond the scope of the present paper.  Nevertheless, the hybrid chaotic inflationary model we present is easily generalized to supersymmetric Pati-Salam or an SO(10) GUT model (see for example, Refs. \cite{Senoguz:2003hc,Antusch:2010va}).

\section{Pati-Salam Generalization}

Consider the following generalization of the above model to $SU(4)_C \times SU(2)_L \times SU(2)_R$ gauge symmetry times $\mathbbm{Z}_4^R$ discrete R symmetry. One family of quarks and leptons are found in the fields $ Q = (4, 2, 1,1) \supset \{q, \; \ell \}, \;  Q^c = (\bar 4, 1, \bar 2, 1) \supset \{ (\begin{array}{c}  u^c \\ d^c \end{array}), \; (\begin{array}{c} \nu^c \\ e^c \end{array}) \}.$  The inflaton sector of the theory is unchanged.  However, now the waterfall fields $\{ S_1, \ S_2 \}$ are replaced by the fields $ \{ \chi^c = (\bar 4, 1, \bar 2, 0), \; \bar \chi^c = (4, 1, 2, 0) \} $.  The superpotential is given by \be  \fw_I = \lambda \ X ( \bar \chi^c \chi^c - \frac{v^2_{B-L}}{2})  +  \Phi (g \ \bar \chi^c \chi^c + m \ Y ) + \bar \chi^c \ \Sigma \ \bar \chi^c +  \chi^c \ \Sigma \ \chi^c \ee where the superfield, $\Sigma = (6,1,1,2)$, is needed to guarantee that the effective low energy theory below the PS breaking scale is just the MSSM.  Note, with the given particle spectrum and $\mathbbm{Z}_4^R$ charges, we have the following anomaly coefficients, \be A_{SU(4)_C-SU(4)_C-\mathbbm{Z}_4^R} = A_{SU(2)_L-SU(2)_L-\mathbbm{Z}_4^R} = A_{SU(2)_R-SU(2)_R-\mathbbm{Z}_4^R} = 1 ({\rm mod}(2)) . \ee Thus the $\mathbbm{Z}_4^R$ anomaly can, in principle, be canceled via the Green-Schwarz mechanism, as discussed in Ref. \cite{Lee:2010gv,Lee:2011dya}.  Dynamical breaking of the $\mathbbm{Z}_4^R$ symmetry would then preserve an exact R parity.

At the end of inflation, $\chi^c, \ \bar \chi^c$ obtain VEVs in the right-handed neutrino directions, spontaneously breaking $SU(4)_C \times SU(2)_L \times SU(2)_R \times \mathbbm{Z}_4^R$ to $SU(3)_C \times SU(2)_L \times U(1)_Y \times \mathbbm{Z}_4^R$.   The superpotential is given by $\fw_I + \fw_{PS}$ with  \be \fw_{PS} =   \frac{\lambda_i^2}{2 M_i} (\bar \chi^c \ Q^c_i)^2 + \lambda Q_3 \ {\cal H} \ Q^c_3 + {\rm higher \; order \; terms} - \mu {\cal H}^2 \ee  where  $ Q_i = (4, 2, 1,1), \;  Q^c_i = (\bar 4, 1, \bar 2, 1)$ with $i = 1,2,3$, a family index and ${\cal H} = (1, 2, \bar 2, 0)$.  This superpotential leads to top-bottom-$\tau$-$\nu_\tau$ Yukawa unification at the PS breaking scale.  If we assume that the Pati-Salam symmetry is the 4D effective theory of an orbifold GUT in higher dimensions, then we also have approximate gauge coupling unification at the PS breaking scale.  In a future paper we propose to write down a complete three family Pati-Salam model with a supersymmetry breaking sector (see, for example, Ref. \cite{Dermisek:2005sw} and subsequent analyses, Ref. \cite{Anandakrishnan:2012tj,Anandakrishnan:2013cwa}).  In this theory we can then discuss leptogenesis and dark matter.

\section{Summary}

In this paper we have presented a novel formulation of chaotic hybrid inflation in a supergravity model with gauged $U_1(B-L)$.   The present model
is an excellent starting point for constructing SUSY GUT models of inflation, including leptogenesis via right-handed neutrino decays.  We have not
considered supersymmetry breaking and issues concerning dark matter, the gravitino or moduli problems in this paper.  We save this for future model building.  Our focus in this paper is proving the compatibility of this new formulation of chaotic, hybrid inflation with recent results from Planck, WMAP and Bicep2.  We find the tensor to scalar ratio $r = 0.163 \ ( 0.148 ) \ [ 0.136 ])$ and the spectral index $n_S = 0.959 \ ( 0.963 ) \ [ 0.966 ]$  for $N_e = 50 \ (55) \ [60]$.
In addition the energy density of the inflaton field during inflation is given by $[V(\Phi_*)]^{1/4} = M_{GUT} = 2 \times 10^{16}$ GeV.   At the end of inflation the waterfall fields oscillate around their minima, spontaneously breaking gauged $B-L$.  Finally reheating occurs when the waterfall fields decay into the lightest right-handed neutrino with an assumed mass of order $10^{10}$ GeV, resulting in a reheat temperature, $T_{RH}$, of order $10^9$ GeV.

As a final note,  this particular model has two issues which can be resolved without much work.  When the $U(1)_{B-L}$ symmetry is broken by the waterfall field, cosmic strings will generically be produced.  Cosmological problems with such cosmic strings can be suppressed by requiring $v_{B-L} \sim 5 \times 10^{15}$ GeV \cite{Buchmuller:2012wn}.   The model as it stands also has kinetic mixing between the B-L and hypercharge gauge bosons.   Both of these issues can be resolved
by starting with the gauge symmetry $SU(3) \times SU(2)_L \times SU(2)_R \times U(1)_{B-L}$.   With a waterfall field which breaks $SU(2)_R \times U(1)_{B-L}$ to $U(1)_Y$ at the end of inflation,  both issues are resolved.  There are no cosmic strings produced \cite{James:1992wb} and there is no kinetic mixing.

\vspace{.5 in}

{\bf Acknowledgements}

{\small S.R. is partially supported by DOE grant DOE/ER/01545-903.
We are also grateful to Archana Anandakrishnan, Charles Bryant, Jessica Goodman and Zijie Poh for discussions.}


\bibliography{bibliography}

\bibliographystyle{utphys}

\end{document}